# Multimedia Content Distribution in Hybrid Wireless Networks using Weighted Clustering

Adrian Andronache
University of Luxembourg
Faculty of Science, Technology and Communication (FSTC)
6, rue Richard Coudenhove-Kalergi
L-1359 Luxembourg

adrian.andronache@uni.lu

Matthias R. Brust
University of Luxembourg
Faculty of Science, Technology and Communication (FSTC)
6, rue Richard Coudenhove-Kalergi
L-1359 Luxembourg

matthias.brust@uni.lu

Steffen Rothkugel
University of Luxembourg
Faculty of Science, Technology and Communication (FSTC)
6, rue Richard Coudenhove-Kalergi
L-1359 Luxembourg

steffen.rothkugel@uni.lu

## ABSTRACT
Fixed infrastructured networks naturally support centralized approaches for group management and information provisioning. Contrary to infrastructured networks, in multi-hop ad-hoc networks each node acts as a router as well as sender and receiver. Some applications, however, requires hierarchical arrangements that—for practical reasons—has to be done locally and self-organized. An additional challenge is to deal with mobility that causes permanent network partitioning and re-organizations. Technically, these problems can be tackled by providing additional uplinks to a backbone network, which can be used to access resources in the Internet as well as to inter-link multiple ad-hoc network partitions, creating a hybrid wireless network. In this paper, we present a prototypically implemented hybrid wireless network system optimized for multimedia content distribution. To efficiently manage the ad-hoc communicating devices a weighted clustering algorithm is introduced. The proposed localized algorithm deals with mobility, but does not require geographical information or distances.

## Categories and Subject Descriptors
C.2.1 [**Network Architecture and Design**]: *Distributed networks, network communications, network topology, wireless communication.*

## 1. INTRODUCTION
Multi-hop ad-hoc networks are composed of a collection of devices that communicate with each other over a wireless medium [6]. Such a network can be formed spontaneously whenever devices are in transmission range. Joining and leaving of nodes occurs dynamically, particularly when dealing with mobility in ad-hoc networks. Potential applications of such networks can be found in traffic scenarios, environmental observations, ubiquitous Internet access, and in search and rescue scenarios as described in detail in [18].

Ad-hoc networks emphasize flexibility and survivability of the whole system. However, centralized approaches e.g. for group management and information provisioning do not work well in such settings. Moreover, due to frequent topology changes, connectivity of devices cannot be generally guaranteed. In particular, this makes it hard to disseminate information in a reliable way.

We overcome these limitations inherent to pure ad-hoc networks by (a) establishing local groups of communicating devices in a self-organizing manner and (b) introducing dedicated uplinks to a backbone infrastructure. Such uplinks are used for accessing resources available in the Internet. Additionally, they are employed to directly interconnect distant devices, either within a single partition as well as across different partitions. In practice, uplinks are realized for instance using cellular networks, satellites, or via Wi-Fi hotspots [6]. Hence, ad-hoc networks with devices that provide uplinks are called hybrid wireless networks throughout this paper. Note that uplinks normally imply additional costs and obey lower bandwidth, so that the uplink has to be applied cautiously.

We show that the proposed clustering mechanism fosters efficient information dissemination within the ad-hoc neighborhood as well as to limit the use of uplinks. The clustering mechanism proposed works locally and is scalable. Additionally, a heuristic weight function is presented for different kinds of device characteristics as well as for topological attributes.

Section 3 describes a prototypical application called HyMN. In Section 4 the proposed clustering mechanism, namely WACA (Weighted Application Aware Clustering Algorithm), and the heuristic weight function is described in detail. An empirical study of WACA is conducted in Section 5. The paper concludes with a discussion of results and future work.

## 2. RELATED WORK
## 2.1 Hybrid Wireless Networks

Infrastructured wireless networks are mainly limited in bandwidth and connectivity. A hybrid setting, such as a hybrid wireless network, is a viable networking solution to tackle these problems. There are a wide range of approaches to establish hybrid networks that connect different communication technologies.

The iCAR approach [22] is built upon the view to support cellular base stations by introducing additional relay stations that work in ad-hoc mode. For this, these relay stations are equipped with two interfaces, one for ad-hoc communication (WLAN) and one for communication with base stations. The major objective of iCAR is to balance the traffic load between cells. Load is forwarded to free cells via ad-hoc relay stations. Additionally, ad-hoc relay stations provide iCAR with increased coverage area.

UCAN [14] also combines cellular and ad-hoc networks (CDMA/HDR and IEEE 802.11b). UCAN uses multi-hop routing to improve the throughput. In contrary to iCAR it is assumed that nodes are completely under the coverage of one base station.

As UCAN, the Hybrid Wireless Network (HWN) architecture [13] tries to optimize the throughput. Thereby HWN requires geographical positions of mobile devices in order to decide if a cell has to be managed in a single-hop or multi-hop ad-hoc mode.

A-GSM [1] focuses on providing connectivity to dead spot areas for ad-hoc networks. Devices are equipped with both, a GSM interface and an ad-hoc communication interface. While one interface is working, the other one is able to check for an alternative connectivity mode.

Ratanchandani et al. [16] focuses on hybrid networks with Mobile IP capability and Internet Gateways to communicate with wired correspondent nodes. Andreadis [2] describes a similar arrangement. It is used a fixed Internet Gateway as for example an access point to provide Internet connectivity to the entire mobile ad-hoc network.

The hybrid setting of Sun et al. [19] consists of base stations that are inter-connected and mobile devices that can connect locally via an ad-hoc mode or to a base station if near enough to it. Two routing schemas are introduced to deal with different application requirements. Sun et al. research points out that the efficiency of the chosen communication mode strongly depends on the applications running in the overall network.

Fujiware et al. [9] proposes a routing protocol and MAC protocol for a multi-hop hybrid wireless network in case of emergency communication. In [3] further protocols for hybrid networks are presented and discussed.

He et al. [11] proposes a centralized peer-to-peer video streaming over hybrid wireless network. The approach is to deliver the base layer of the video from the server via WLAN, while the enhancement layers uses multiple paths in ad-hoc mode.

None of the mentioned approaches and architectures here deals with chunking issues of the requested information or parallel injection of information to selected devices.

## 2.2 Clustering and Clusterhead Selection

Clustering algorithms can be based on criteria such as energy level of nodes, their position, degree, speed and direction. Centralized and distributed approaches can be distinguished, as well as probabilistic and deterministic ones. Probably the most crucial point when dealing with clustering is the criterion how to choose the clusterhead. The number of clusterheads strongly influences the communication overhead, latency, inter- and intra-cluster communication design as well as the update policy (i.e. execution of re-organization of clusters).

One of the first and most influential cluster-based protocols is LEACH [12]. It uses a distributed probabilistic approach. Each node elects itself as a clusterhead with a certain probability based on the desired percentage of the clusterheads in the network, and the last round where it served as a clusterhead. Thus, the role of the clusterhead is probabilistically rotated, which enables to save a large amount of energy.

In [15], a centralized clusterhead election algorithm is presented, where the base station assigns the clusterhead roles based on the energy level and the geographical position of the nodes.

In [10], a centralized algorithm based on fuzzy logic is proposed. The nodes are selected as clusterheads by the base station based on their distances to each other, energy level, and the concentration of the nodes in the region.

Chatterjee et al. [4] propose a distributed deterministic clusterhead selection algorithm, namely WCA (Weighted Clustering Algorithm). For reasons that the proposed WACA clustering algorithm is compared to WCA in this paper, WCA is described in more detail here.

WCA obtains 1-hop clusters with one clusterhead. The election of the clusterhead is based on the weight of each node. For this a heuristic weight function is used that uses distances between the neighbors, degree (number of neighbors), speed of neighboring nodes, and battery power of the node as well as weighting factors to calculate the weight. To obtain this information, WCA assumes to be provided with geographical information or relative distances of one node and its surrounding. The WCA update policy is triggered to be invoked by isolated nodes on demand. Special cases are detachment of current clusterhead and attachment to a new clusterhead. The clusterhead continuously sends a message to its neighbors. The neighbors check if the signal strength decreases what implies that the distance to the clusterhead is increasing. In that case, the node informs its current clusterhead that it detaches and chooses the next available clusterhead. If there is no clusterhead available, the election procedure is evoked to create a new cluster. Observe that the continuous message exchange is a principal drawback of that algorithm.

Tan et al. [20] present a distributed clusterhead selection algorithm where each node computes its priority based on its ID, current communication round, energy level and speed. This information is exchanged within the two-hop neighborhood. The nodes with highest priority become clusterheads.

Early approaches as [7] describe an ID-based clusterhead selection algorithm. Each node in the network is assigned a unique ID. The selection process consists in designating locally the device with the lowest ID as clusterhead. No further parameters are used in this approach.

None of the algorithms introduced before gives guarantees on the resulting network structure, e.g. on the number of the resulting clusterheads. Their effectiveness is evaluated by simulation. In this sense, the aforementioned algorithms realize heuristics. Our approach also falls into this category.

## 3. HYMN

In this section we describe a hybrid wireless network model together with a prototypical application called HyMN. The HyMN (Hybrid Multimedia Network) application aims at sharing multimedia files efficiently.

### 3.1 Hybrid Wireless Network Model

We focus on establishing clusters in a self-organized way. For this, a clustering algorithm, namely WACA, which uses a heuristic weight function, is proposed. The WACA algorithm is designed to build an ad-hoc network topology that fits the needs of the application running on top of it. To achieve this, several parameters can be set in the weight function. For instance if an application requires much CPU and memory usage for the clusterheads, the CPU and memory load of the devices will play a central role in the weight calculating process. The HyMN application focuses on multimedia content distribution from a backbone network to ad-hoc networks. To optimize this process, parameters like signal strength to the backbone network, long battery lifetime, dissemination degree and clustering coefficient are important. Details about the WACA system parameters, used in the weight calculation function of the HyMN application are presented in Section 4.2.

Let's assume a considerable number of mobile devices. Each device is able to use radio technology like for instance Bluetooth or Wi-Fi for wireless communication in the physical proximity. The devices are considered to be heterogeneous, since each of them might be supplied with different memory capacities, computational power, and available energy. Some of these devices are supposed to be equipped with uplink-capable adapters such as GSM, 3G or satellite.

Locally, the ad-hoc network devices elect a clusterhead that is in charge of keeping track of local devices and their shared interests. Clusterheads are chosen according to their weight that is calculated by a heuristic weight function (cf. section 4.3). Parameters like available power, signal strength, topological position etc. are taken into account by means of this function.

Clusterheads might also act as injection points [17]. Injection points maintain a connection to the backbone network and request information related to the common *interests* shared by the devices in the cluster. The term interest in this respect might refer for example to news, weather information, certain game high score tables and more.

As soon as the backbone can provide fresh information related to a particular interest, it is injected into the multi-hop ad-hoc network by sending it to the injection point. Updates of information are also injected as long as the injection point keeps the connection to the backbone.

The injection point disseminates the injected information over a wireless connection like Wi-Fi or Bluetooth to the interested devices, which form a logical interest group. These connections have the advantage to be free of charge. Moreover, these technologies allow higher bandwidth compared to GSM or 3G cellular connections. In a scenario where a cellular flat rate is not available, the injection point is finally also in charge of splitting the cellular connection costs to devices receiving the information. Splitting of costs is an issue of further investigations.

### 3.2 The HyMN Application

The HyMN application is designed for users interested in live multimedia news from a certain sports event such as the Football World Championship. For instance football fans have the possibility to create an interest group in a local ad-hoc network partition. Examples scenarios include football fans in pubs, those watching another match, traveling ones, and more. In each of these cases, a considerable number of people have shared interest and might join forces in a local setting.

Interest in certain football matches is registered with the clusterhead. The clusterhead maintains an uplink to the backbone network—acting as injection point—in order to receive multimedia news related to the interests of the ad-hoc members. Thus, the football fans will receive injected information such as small videos, pictures or text messages each time something interesting is happening during the game. The multimedia files received remain stored on the mobile devices and will be provided to further interested devices in the ad-hoc network.

Instantiating the idea introduced more generally above, the devices in the ad-hoc network running HyMN elect clusterheads that are in charge of registering with the backbone using the cellular or satellite network, thus acting as injection points. The backbone injects a list of available football matches to the injection points, which provide it to the ad-hoc network devices. The list is injected each time it is updated, e.g. when football matches end or new ones begin.

After receiving the list, the HyMN devices are able to choose their interests, i.e. from which game they want to receive text news, images or short videos. The injection points are in charge of keeping track of local interests and to register them with the backbone, which injects appropriate files as soon as available.

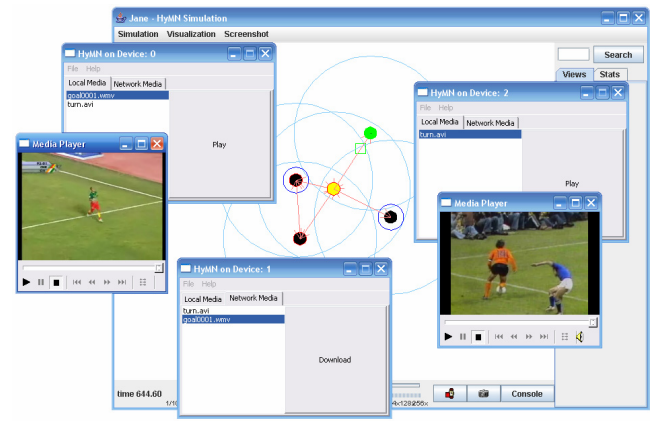

**Figure 1. Prototypical implementation of the HyMN application. Different users are sharing media files.**

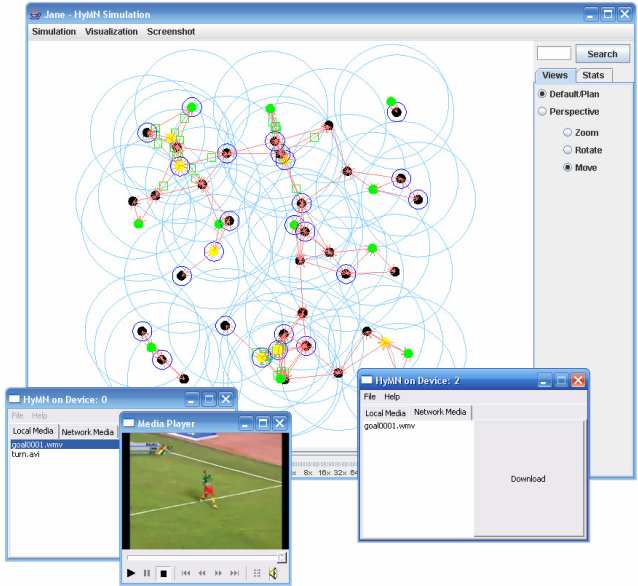

**Figure 2. The HyMN application establishing the WACA network topology.**

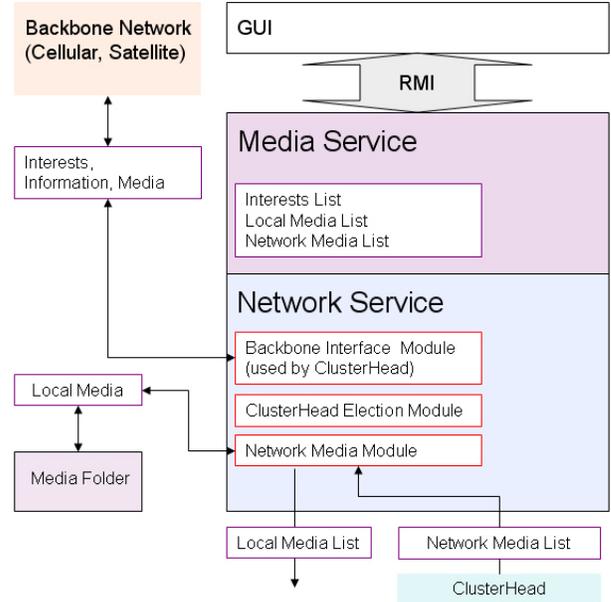

**Figure 3. The HyMN architecture**

The backbone splits the multimedia files into chunks, which will be injected concurrently to different injection points belonging to the same ad-hoc network partition. Thus, the overall bandwidth for injecting data is increased and the files reach the ad-hoc network faster. After reaching the network, the chunks are exchanged among the interested devices until all devices receive the complete file.

The HyMN application has been implemented prototypically on top of the JANE Simulator [8] (Figure 1). JANE is designed to support application and protocol design in the realm of ad-hoc networks. One distinct feature of JANE is that applications can be executed in a simulation mode and in a hybrid mode with real devices being attached to a running simulation.

## 3.3 The HyMN Architecture

The prototypical implementation of the HyMN application consists of two main services: *network service* and *media service* (Figure 3).

The *media service* encompasses the graphical user interface (GUI) that provides information about local media, network media and interests offered. This service sends the user requests to the *network service* using the JANE signal mechanism. Communication with the GUI is done using the Java Remote Method Invocation (RMI). Thus, the GUI can be started on demand either on the simulation machine or on a mobile device if the simulation is running in hybrid mode.

The *network service* provides mechanisms for local network communication, using the Bluetooth or Wi-Fi interface, and for backbone communication, using a 3G or satellite interface. This service contains also two important modules: *clusterhead election module* and *network media module*.

The *clusterhead election module* is in charge of reacting on network changes and to decide which neighbor is the responsible clusterhead based on local information (cf. Section 4). The *network media module* keeps track of media files available on the own device and exchanges the list with the neighbor devices. If updates arrive, lists containing the own and the known network media files are sent to the *media service* in order to update the graphical user interface (Figure 2). This module can be requested to download media files from neighbors or to inform the clusterhead about own interests.

## 4. WACA

In this Section we present the WACA algorithm that aims to elect local clusterheads in ad-hoc networks and to create clusters in order to build a beneficial topology for the hybrid network application HyMN. The goal is to provide a simple, scalable and power aware algorithm in order to build a topology that promotes one-hop communication and aims at reducing multi-hop communication. This way, the overall network load is decreased while failure resilience is optimized.

## 4.1 Algorithm

One objective of WACA is to avoid network communication overhead during the clusterhead election and clustering process. Therefore, the election of a local clusterhead is based solely on information available locally. To achieve this, each device calculates its own weight based on its device parameters like remaining power, backbone signal strength and topological attributes (cf. Section 4.2). The weight is recalculated when changes of attributes occur. Each device propagates its own weight as part of the beacon, which is a periodically broadcasted message used in ad-hoc networks to detect devices in communication range. The algorithm only considers so-called neighbor devices, i.e. devices that mutually see each other.

The pseudo code of the WACA algorithm is shown in Fig. 4. The devices run the algorithm each time the set of neighbor devices changes, e.g. when devices enter or leave the communication range, or when weights are updated. Using the information about the neighborhood, each device elects the neighbor device with the highest weight value as clusterhead. Devices that use a mutual clusterhead are called *cluster slaves* of that device.

**Algorithm WACA** for device $d$
**Input :** $A$: Set of Device attributes
$N$: Set of neighbor devices
$w: N \mapsto \mathbb{N}$, weights for elements in $N$
$c: N \mapsto N$, current cluserhead
**Output :** $w(d)$: Weight of device $d$
$c(d)$: Cluster head of device $d$

1: $isSubHead \leftarrow false$
2: $isClusterHead \leftarrow true$
3: $c(d) = d$
4: $w(d) \leftarrow calculateWeight(A)$
5: **for each** $n \in N$ **do**
6:   **if** $w(n) > w(c(d))$ **then**
7:     $c(d) \leftarrow n$
9:   **if** $c(n) = d$ **then**
10:     $isSubHead \leftarrow$ **true**
13: $setBeaconData(w(d), c(d))$

**Figure 4. Pseudo code of the WACA clustering algorithm.**

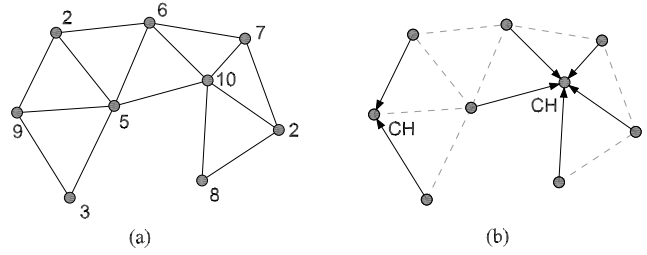

**Figure 5. Topology before (a) and after (b) applying WACA. Two clusterheads (CH) are established.**

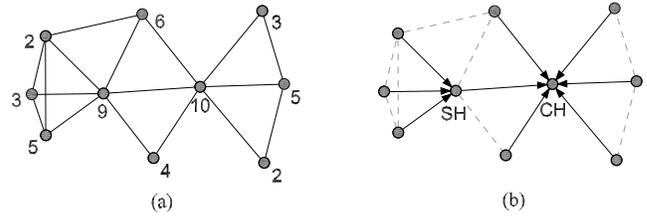

**Figure 6. Topology before (a) and after (b) applying WACA. The resulting topology includes one clusterhead (CH) and one sub-head (SH).**

Clusters are created in a hierarchically fashion. Each device elects exactly one device as its clusterhead, i.e. the neighbor with the highest weight (Figure 5). This clusterhead in turn also investigates its one-hop neighborhood, also electing the device with the highest weight as its clusterhead. This process terminates in case of a device electing itself as its own clusterhead, due to the fact of having the highest weight among all its neighbors. We call all intermediary devices along such clusterhead chains *sub-heads* (Figure 6). Each device on top of a chain is called a full clusterhead, or, in short, just clusterhead. Hence, in each network partition, multiple clusterheads might coexist. Only (full) clusterheads might serve as injection points, depending on current interests of the cluster members. Information will be injected to clusterheads only. The sub-heads in turn are responsible for further forwarding the data to other cluster members, i.e. its cluster slaves. Some of them might act as sub-heads, too.

As mentioned above, a sub-head will be in charge of the same tasks as a clusterhead concerning its own cluster slaves. The only difference to a clusterhead is the fact that the sub-head will not maintain a backbone connection and will forward the requests and information from its cluster slaves to its own clusterhead. In case that a sub-head is loosing the connection to its clusterhead, it will connect to the backbone, thus becoming a clusterhead itself. Of course, timing issues are of importance in this respect and will be discussed further in Section 4.4.

### 4.2 System Parameters
Appropriate selection of parameters for calculating weights is a crucial point. Our approach focuses on both, augmenting stability of the clustering topology and fulfilling the needs of the HyMN application. For this, we chose as parameters device power, signal strength, clustering characteristics, and any changes of those crucial parameters in neighborhood.

#### 4.2.1 Device Power
Nowadays, wireless connections like Wi-Fi or Bluetooth are providing a considerable higher bandwidth than 3G cellular connections. Furthermore, 3G communications also consumes more energy. By definition, clusterheads that also act as injection points rely on both types of connections, both locally as well as for the uplink. They need to keep track of local administration and to provide injected information to interested devices in the own cluster. Hence, the remaining battery power of devices is to consider when electing a clusterhead. Moreover, in view of our approach a device appears to be more appropriate in particular when the available power is higher than that of the other devices in the neighborhood.

#### 4.2.2 Signal Strength for Backbone Connectivity
In case of cellular networks, devices closer to a 3G CDMA base station perceive higher data rates than devices farther from the base station. A higher distance in turn typically results in intermittent connectivity and lower data rates. Electing devices closer to the base station provides higher data rates for data injection [14]. Once the data has reached the ad-hoc network, it can be disseminated via high bandwidth links, e.g. Wi-Fi, to all devices in that injected cluster. The influence of the quality of the uplink is much higher than that of the local ad-hoc communication. That's why we explicitly omitted the ad-hoc signal strength from selecting proper clusterheads.

### 4.2.3 Dissemination Degree

A clusterhead is a pivot entity for information distribution. Information requests, cluster management, backbone-driven information injection and also distribution of that information to interested devices are the jobs of a clusterhead that have to be scheduled appropriately. To disseminate injected information efficiently a clusterhead should have a higher degree than its neighbors, i.e. more connections to other devices than each device in the communication range. In practice the ideal degree of a device depends strongly of the technology employed, as pointed out in [5]. For instance the master-slave model used in *Bluetooth* handles up to seven slaves. In that case, a higher degree than seven causes latency in information delivery. Like in [5], we introduce a parameter $dd_I$ that represents the ideal degree for a clusterhead to achieve a high throughput.

### 4.2.4 Local Clustering

Due to the introduction of sub-heads, WACA creates multi-hop clusters of undetermined size. Applying WACA in large network partitions might result in a high number of hops to the clusterhead. This is due to the possibly long chain of sub-heads attached to a clusterhead. To tackle this problem locally, we propose the use of a local clustering coefficient as described in [21] to support well-structured clusters. Informally speaking, the local clustering coefficient $c_L$ is capturing the connectivity between the neighbors of one node. For details cf. Section 4.3. Additionally, the local clustering coefficient can help in identifying topological unfavorable devices. For instance devices close to partition borders can be assumed to leave the partition earlier than more centralized ones. The local clustering coefficient is a local aid to limit the cluster size taking scalability of WACA into account.

### 4.2.5 Stability Coefficient

Due to mobility as well as network activity the cluster characteristics change over time. Depending on the update policy, the validity of clusterheads has to be checked and if necessary a clusterhead re-election has to be invoked. Cluster re-organization, however, causes additional message exchanges and computational complexity. In situations where few parameters of the clusters changed—e.g. one new neighbor is recognized or a current neighbor leaves—it is reasonable to keep the existing clusterhead instead of re-electing a new one. Moreover, in cases where e.g. HyMN already started an injection, it is important not to interrupt this process. Therefore, we introduce a stability coefficient *s* that assigns an additional weight to the current clusterhead, increasing the probability of keeping it. For example, if a clusterhead is surrounded by a group of devices and they are moving together in the same direction, it could happen that new neighbors are discovered from time to time. Since they disappear from the neighbor list after the group has passed, it is beneficial not to re-organize the cluster just for those minimal and temporary changes, even if the new device is of considerable higher weight.

## 4.3 Heuristic Weight Function

In this Section, the system parameters are combined with certain weighing factors. The flexibility of changing the weighing factors allows applying the algorithms for very different networks as well as applications. Due to the fact that the network topology is built based on the weight of the devices, the weight calculation plays a central role in our algorithm. To determine the weight, following points have to be performed. Hereby we assume that the neighbor discovery service already filled the neighbors list *N(d)* on one device *d* with the IDs of those devices within the transmission range of *d*. *D* represents the set of devices.

$$N(d) = \bigcup_{d' \in D,\, d' \neq d} \{dist(d,d') < r\}. \qquad (1)$$

**Device power.** Given a device *d* with available power *P(d)*, then calculate the power-appropriateness $P_A$ of device *d* as

$$P_A = \frac{3}{2} + \frac{1}{2}\log\left(P(d) - \frac{3}{5}\right). \qquad (2)$$

**Signal strength.** Usually the strength of the backbone network signal is available on each device, e.g. the signal strength to a cellular network base station. Let *s* be the strength of the signal, given by a value between 0 and 1.

**Dissemination degree.** Compute the difference between ideal degree $dd_I$ for device *d* and real degree $|N(d)|$ (cf. [5]) as

$$\Delta dd = 1 - \frac{|N(d) - dd_I|}{dd_I}. \qquad (3)$$

**Local clustering coefficient.** Compute the local clustering coefficient $c_L$ of one device *d* that is defined as

$$c_L = \frac{|N(d)|}{n(n-1)/2}, \qquad (4)$$

where $|N(d)|$ is the number of links in the neighborhood of *d* and $n(n-1)/2$ is the number of all possible links, whereby *n* is the number of all devices.

**Stability coefficient.** Calculate the stability-difference between the neighborhood *N'(d)* at the time when *d* was elected with the current neighborhood *N(d)* as

$$\Delta N = 1 - \frac{|(N'(d) \setminus N(d)) \cup (N(d) \setminus N'(d))|}{|N(d)| + |N'(d)|}. \qquad (5)$$

In case that device *d* is currently not the clusterhead, assign value 0 to $\Delta N$.

**Calculating weight function.** Calculate the total weight of a device *d* as

$$W_d = wf_1 P_A + wf_2 s + wf_3 c_L + wf_4 \Delta dd + wf_5 \Delta N, \qquad (6)$$

where $wf_1$, $wf_2$, $wf_3$, $wf_4$, $wf_5$ are weighing factors choosing according requirements.

Choosing a clusterhead depends on the parameters described and the related weighing factors. Due to the different performance of the mobile device batteries the energy level of the batteries cannot be used as metric for *P*. For instance a notebook with 70% battery level will mostly outlive in the injection point role a PDA that has a 100% battery level. Thus, parameter $P_A$ represents the power-appropriateness being a clusterhead for the designed job, e.g. receiving a football clip of certain size. Creating a weight of the available power on one device, we have to take in account that it is a extremely inappropriate situation when the device owns of just e.g. 60 % of the estimated required power for the job as injection point.

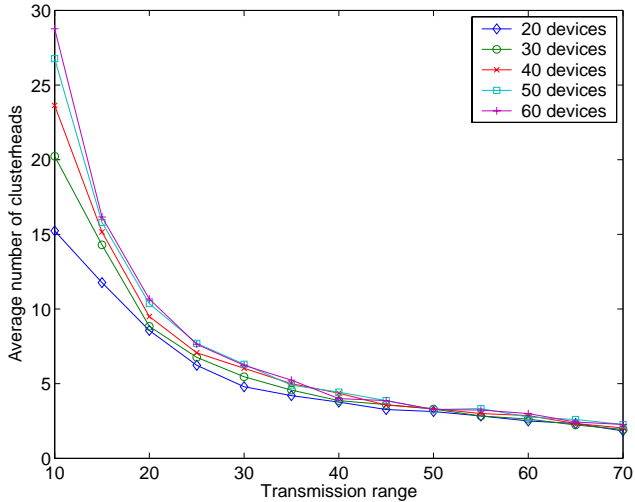

**Figure 7. Average number of clusterheads.**

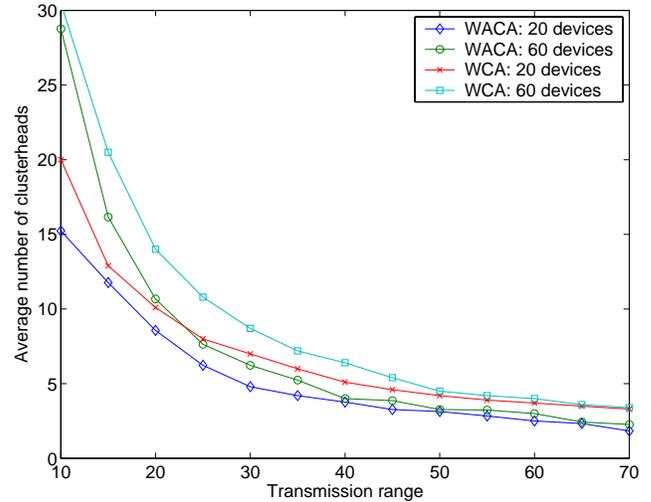

**Figure 7. Comparing the average number of clusterheads of WACA and WCA.**

On the other side, having a lot more as 100 %, e.g. 400 %, of that power should not influence too much the weight not for the other parameters being underrated. For this reason, we choose a log-function. A power-appropriate device may suffer low bandwidth due to bad connectivity with the base station. The current signal strength is assigned to parameter *s* as described above. The MAC layer supports an ideal number of devices in its neighborhood that it can best manage. Often, there are less or more neighbors. The clusterhead election procedure also elects a device as clusterhead according to how near the current number of neighbors is to that ideal number. The parameter $\Delta dd$ reflects the difference between real and ideal degree (i.e. number of neighbors). The closer this value is to 1, the better it fits with the ideal number of neighbors. As pointed out, the sub-head introduction may cause unsuitable cluster structures. The local clustering coefficient weighs balanced clusters higher. Implied by the clustering coefficient, clusterheads tend to be at the center of a group of devices. Hence, the effect of mobility is kept low with respect to the clusterhead and the resulting topology is more stable. Due to the mobility of nodes it might happen that the local neighborhood changes very fast, e.g. a fast node passes closely alongside the clusterhead. In such cases it is typically more beneficial to stick with the current clusterhead instead of changing it temporarily only. Formula (5) expresses the changes that happened in the neighborhood of a device *d*. That can be new devices that appeared and already known devices that left. A value close to 0 indicates significant changes in the neighborhood of *d*.

### 4.4 Update Policy and Message Complexity

When the ad-hoc network is initially established, each device calculates the own weight and disseminates it through beaconing in the neighborhood as described in Section 4.1. After receiving the beacon from all neighbors, a device will elect the neighbor with the highest weight as its clusterhead. During the lifetime of the ad-hoc network, re-election of clusterheads can occur, due to possible changes of parameters in the neighborhood. The devices are tracking changes of the own parameters, recalculate the own weight and update the beacon. Thus, the devices are always up to date with respect to the current weight of their neighbors.

In case the clusterhead vanishes, e.g. moves out of the cluster's communication range or is switched off, each cluster slave will check its neighbor list and elect the one with the highest weight as new clusterhead. If the system parameters of the clusterhead change so that the device cannot accomplish the injection point tasks anymore, then the weight will be drastically lowered, thus inducing a re-election in the neighborhood. There may be situations where two moving clusters meet. In this case, the neighborhood of both clusterheads changes. This will lead to a recalculation of the stability coefficient on both devices, and thus the weights are updated accordingly.

The WACA algorithm calculates a total weight on each device. After exchanging these weights to the neighbors the device with the highest weight is considered to be a clusterhead. Since each device has to beacon the weight to its neighbors, the message complexity is $O(n)$, where *n* are the number of mobile devices in the network.

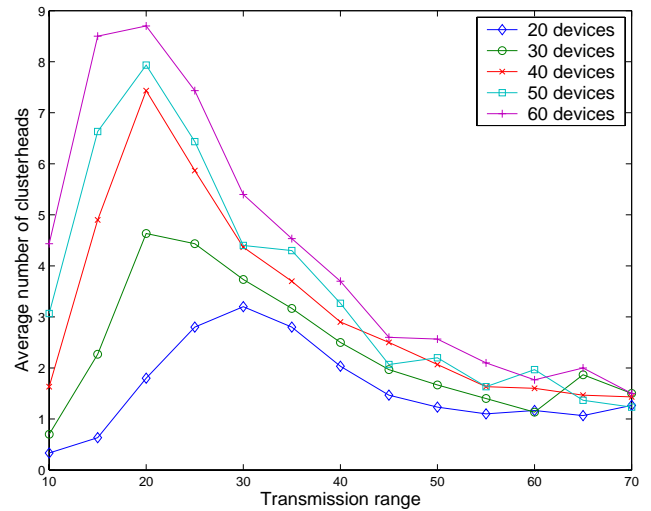

**Figure 8. The average number of sub-heads for different number of devices using WACA.**

## 5. EMPIRICAL STUDY

We choose a 100×100 unit square as basic simulation setting. A number of $N$ nodes are deployed uniformly by random using a validated random number generator initialized by independent seeds. In the simulation the number was set to values between 20 and 60. The transmission range varied between 10 and 70 with a fixed step of 5. The weighing values were in accordance to the requirements of the HyMN application and set to $wf_1 = 0.9$, $wf_2 = 1$, $wf_3 = 0.85$, $wf_4 = 0.65$, $wf_5 = 0.6$. Observe that for different application requirements, the weighing factors have to be adjusted. All results are averaged over 30 simulation runs.

The first experiment provides a detailed report of the average number of clusterheads or clusters, respectively, for the different values of $N$ (Figure 7). For all values chosen for $N$ the average number of clusterheads decreases when increasing the transmission range. We argue that this asymptotic behavior is because a clusterhead with larger sending range covers an exponentially larger area.

In a second experiment we compared the average number of clusters of WACA with that of WCA [5]. For this, $N$ is set to 20 and 60 using both algorithms. The transmission range is varied as described above. The results show that the average number of clusterheads using WACA in both cases is below of that of WCA (Figure 7). Not illustrated, but shown through further simulation, this also holds for the cases where $N = 30, 40, 50$. As mentioned above, reducing the number of clusterheads strongly influences the communication overhead, latency, inter- and intra-cluster communication design as well as execution of re-organization of clusters. We understand this performance improvement due to the fact that sub-heads have been introduced to allow multi-hop clusters, but keeping the clusters well-formed.

In the next simulation we investigate the number of sub-heads for the same setting of the first experiment. Results are reported in detail in (Figure 8). The number of sub-heads increases as the transmission range increases, and reaches a peak when transmission range is between 20 and 30. Further increase of the transmission range results in a decrease of the average number of sub-heads. This behavior can be explained by the fact that sub-heads cannot be easily established in cases where the transmission range is very low, because clusters tend to be one-hop structures. When increasing the transmission range clusters are getting bigger, encompassing more devices, and an increasing number of sub-heads are established. Further increasing the transmission range results in fewer sub-heads, because the clusterhead can reach more devices directly without the need of an intermediary sub-head. This explains the asymptotic decrease of the average number of sub-heads.

## 6. CONCLUSION AND FUTURE WORK

We introduced a prototypically implemented application called HyMN for multimedia content distribution in hybrid wireless networks. For the management of the ad-hoc nodes, a weighted clustering algorithm (WACA) has been presented.

WACA is explicitly designed for hybrid networks, i.e. the symbiotic combination of multiple ad-hoc network partitions inter-linked by a backbone network. The heuristic weight function design and the device weight calculation are strongly influenced by the requirements of the employing application HyMN.

Results have shown that the average number of clusterheads can be decreased using WACA compared to the WCA clustering algorithm. Note that in contrary to WCA, WACA creates multi-level hierarchical clusters. Additionally, the WACA algorithm does neither depend on geographic positions nor on distances between neighbors, both of which are hard to determine. This makes the implementation of the algorithm on real devices more suitable. WACA works on local information only and supports well-formed multi-hop clusters, realized by introducing cluster sub-heads.

In the future, further experiments have to be performed to verify that WACA spreads information more efficiently than other clustering algorithms as well as that it avoids needless use of uplinks. Furthermore, we will include mobility in our simulation settings. The HyMN application will be further developed and a middleware will be abstracted out of it.